# NeuroSleep: Neuromorphic Event-Driven Single-Channel EEG Sleep Staging for Edge-Efficient Sensing

Boyu Li[1, 2†], Xingchun Zhu[3†] and Yonghui Wu[1*]

[1] School of Flexible Electronics, Henan University, Kaifeng, China
[2] School of Electrical and Electronic Engineering, Nanyang Technological University, Singapore
[3] Division of Life Sciences and Medicine, University of Science and Technology of China, Hefei, China
[†] These authors contributed equally to this work.
[*] E-mail: defey@henu.edu.cn

**Abstract**

*Objective*. Reliable, continuous neural sensing on wearable edge platforms is fundamental to long-term health monitoring; however, for electroencephalography (EEG)-based sleep monitoring, dense high-frequency processing is often computationally prohibitive under tight energy budgets. *Approach*. To address this bottleneck, this paper proposes NeuroSleep, an integrated event-driven sensing and inference system for energy-efficient sleep staging. NeuroSleep first converts raw EEG into complementary multi-scale bipolar event streams using Residual Adaptive Multi-Scale Delta Modulation (R-AMSDM), enabling an explicit fidelity–sparsity trade-off at the sensing front end. Furthermore, NeuroSleep adopts a hierarchical inference architecture that comprises an Event-based Adaptive Multi-scale Response (EAMR) module for local feature extraction, a Local Temporal-Attention Module (LTAM) for context aggregation, and an Epoch-Leaky Integrate-and-Fire (ELIF) module to capture long-term state persistence. *Main results*. Experimental results using subject-independent 5-fold cross-validation on the Sleep-EDF Expanded sleep-cassette (SC) subset with single-channel EEG demonstrate that NeuroSleep achieves a mean accuracy of 74.2% with only 0.932 M parameters while reducing sparsity-adjusted effective operations by approximately 53.6% relative to dense processing. Compared to the representative dense Transformer baseline, NeuroSleep improves accuracy by 7.5% with a 45.8% reduction in computational load. *Significance*. By coupling neuromorphic event encoding with state-aware context modeling, NeuroSleep offers a deployment-oriented framework for single-channel sleep staging that reduces redundant high-rate processing and improves energy scalability for wearable and edge platforms.

## 1. Introduction

Continuous neural sensing on wearable edge platforms is increasingly desired for always-on health monitoring, yet it remains constrained by tight energy and compute budgets [1, 2]. A major bottleneck is that many neural signals, including electroencephalography (EEG), are acquired as dense high-rate time series, while discriminative dynamics are sparse and non-uniform over time [3]. Sleep staging provides a representative long-duration EEG monitoring task where reliable operation requires both cross-subject generalization and long-term temporal consistency under resource constraints [4, 5].

Among neural signals, EEG directly reflects neural activity in the brain and is widely regarded as the most discriminative physiological signal source for sleep staging [6]. However, conventional systems typically rely on multi-channel EEG acquisition and centralized processing, resulting in high system complexity and power consumption that are unsuitable for long term continuous monitoring [7, 8]. In contrast, single-channel EEG offers clear advantages in terms of wearing comfort, system simplicity, and deployment cost, making it more compatible with wearable scenarios [9]. This simplified sensing configuration, however, significantly reduces spatial redundancy and increases sensitivity to noise and inter-subject variability, and makes temporal-context modeling more challenging, thereby imposing stricter requirements on the inference architecture [10, 11].

From a signal-processing perspective, EEG for sleep monitoring is sampled as a continuous time series with high temporal resolution, while the discriminative information for sleep staging is not uniformly distributed over time [12-14]. Large portions of the signal remain relatively stationary and contribute little to sleep stage discrimination, whereas critical cues are mainly concentrated around stage transitions or characteristic physiological events [15]. Nevertheless, most existing automatic sleep staging methods follow a dense processing paradigm, applying feature extraction and inference uniformly across all time segments [16-18]. This leads to substantial computation and memory activity that is weakly coupled to discriminative dynamics, creating a major barrier to energy-efficient edge deployment.

In recent years, neuromorphic computing and spiking neural networks (SNNs) have emerged as a promising approach to address this mismatch [19, 20]. By triggering computation only in response to signal changes rather than continuous sampling, they are inherently well aligned with the non-uniform information distribution of physiological signals [21]. However, existing neuromorphic approaches for EEG analysis often still rely on high-resolution time-step updates or focus primarily on short-range modeling, and thus do not fully meet the practical requirements of sleep staging, including long-term contextual consistency, stable state representation, and continuous low-power operation at the system-level [22-24].

These observations motivate a design principle for edge-efficient sleep staging: high time-resolution processing should be reserved for change-driven information, while long-range sleep context should be modeled on low-frequency abstract representations with explicit continuity control. Such a separation offers a principled pathway to reduce redundant computation and energy consumption without sacrificing staging performance.

Based on this principle, this paper proposes NeuroSleep, an event-driven edge-intelligence system for single-channel EEG sleep staging. NeuroSleep converts raw EEG into complementary multi-scale neuromorphic event streams via Residual Adaptive Multi-Scale Delta Modulation (R-AMSDM), enabling an explicit fidelity–sparsity trade-off at the sensing front end. On top of this event representation, a hierarchical inference architecture is constructed, consisting of an Event-based Adaptive Multi-scale Response (EAMR) module, a Local Temporal-Attention Module (LTAM), and an Epoch-Leaky Integrate-and-Fire (ELIF) state module to support continuity-aware and low-frequency sleep state modeling. By aligning high-resolution processing with change-driven events and long-term context modeling with low-frequency state updates, this design targets robust sleep stage characterization under resource-constrained wearable settings. The overall workflow is illustrated in figure 1.

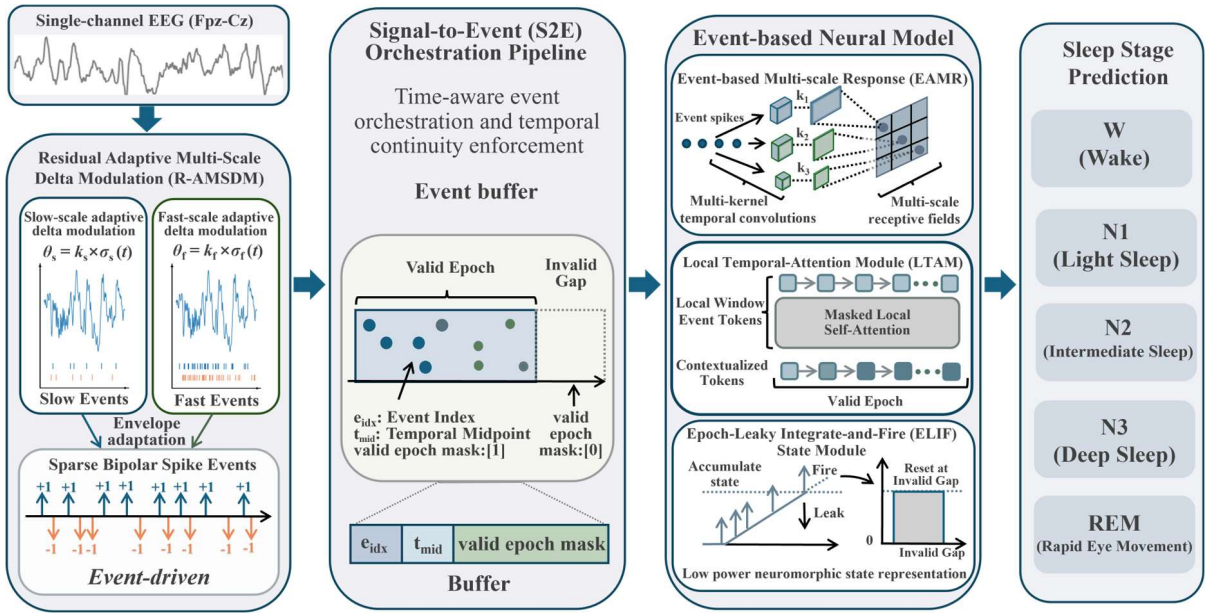

**Figure 1**. Overview of the neuromorphic edge AI pipeline for single-channel EEG sleep staging.

The main contributions of this work are summarized as follows.

1) NeuroSleep is presented as an integrated event-driven neuromorphic sensing-and-inference framework for single-channel EEG sleep staging, in which conventional dense processing pipelines are reformulated from a system-level perspective toward edge-efficient deployment.

2) Residual adaptive multi-scale delta modulation scheme (R-AMSDM) is developed to convert continuous EEG signals into sparse neuromorphic event streams, enabling a controllable trade-off between information fidelity and event sparsity.

3) A hierarchical event-based inference architecture is constructed, incorporating EAMR, LTAM, and an ELIF state module to support continuity-aware and low-frequency sleep state modeling.

4) Comprehensive evaluations are conducted on a public sleep dataset, where both performance-oriented metrics and deployment-relevant indicators, including accuracy, model complexity, latency, computational cost, and event sparsity characteristics, are systematically reported.

## 2. Neuromorphic event encoding

### 2.1 Residual Adaptive Multi-Scale Delta Modulation (R-AMSDM)

In the proposed NeuroSleep system, neuromorphic event encoding serves as the perceptual interface between continuous EEG signals and subsequent event-level inference [25]. Its objective is to preserve dynamic variations relevant to sleep staging while representing the signal as a sparse bipolar event stream, thereby reducing redundant processing of high temporal resolution continuous signals [26, 27]. As illustrated in figure 2(a), a continuous EEG waveform can be equivalently expressed as a sparse sequence of bipolar events, such that computational activity is primarily triggered by segments exhibiting significant signal changes.

Furthermore, discriminative cues in sleep EEG encompass both slowly varying trends and short term fluctuations. As shown in figure 2(b), single scale delta based event encoding tends to generate pronounced regions of redundant triggering, while naive parallel multi-scale encoding often introduces duplicated events across scales. To address this issue, this work proposes Residual Adaptive Multi-Scale Delta Modulation (R-AMSDM), which combines adaptive delta modulation with progressive residual decomposition. Specifically, a low sensitivity scale is first employed to account for dominant trend components, after which a more sensitive event encoding is applied only to the unexplained residual signal. Through this residual driven multi-scale strategy, events generated at different scales become semantically complementary rather than redundant, as illustrated in figure 2(c).

The continuous EEG signal is represented as a discrete time sequence $x(t)$. R-AMSDM triggers events by comparing the current input with a reconstructed reference value. During the encoding process, the system maintains a time-varying reconstruction reference $r(t)$, and computes the differential signal as:

$$d(t) = x(t) - r(t). \tag{1}$$

When the magnitude of the differential signal $d(t)$ exceeds a pre-defined threshold, an event is generated and the reference value is updated accordingly.

To enable adaptive event triggering across different sleep stages, subjects, and temporal activity levels, R-AMSDM introduces a dynamic threshold based on local signal statistics. Specifically, the adaptive threshold $\theta(t)$ is estimated from the local standard deviation of the signal within a causal temporal window $W$:

$$\theta(t) = k \cdot \sigma_{local}(t), \tag{2}$$

where $k$ is a fixed scale factor that controls the sensitivity of event triggering. This design allows the threshold to vary with the local fluctuation level of the signal. In more active segments, a higher threshold is applied to suppress excessive triggering, whereas in relatively stable segments, the threshold is reduced to preserve weak but potentially discriminative signal variations.

At each sampling instant, the event signal $s(t) \in \{-1, 0, +1\}$ is generated by comparing the differential signal with the adaptive threshold, and is used to update the reconstruction reference, given by:

$$s(t) = \begin{cases} +1, d(t) \geq \theta(t) \\ -1, d(t) \leq -\theta(t), \\ 0, \text{otherwise} \end{cases} \quad (3)$$
$$r(t+1) = r(t) + s(t)\theta(t),$$

where the sign of $s(t)$ indicates the event polarity. The reference update forms a closed loop reconstruction mechanism, enabling the encoder to convert the continuous waveform into a sequence of sparse incremental events.

However, single scale delta modulation is insufficient to simultaneously capture the slowly evolving trend components and short term transient fluctuations present in sleep EEG signals. To address this limitation, R-AMSDM further introduces a residual driven multi-scale event encoding mechanism. In this design, two encoding scales with different temporal sensitivities are defined, whose event triggering thresholds are given by:

$$\theta_{\text{slow}}(t) = k_{\text{slow}} \cdot \sigma_{\text{local}}(t), \theta_{\text{fast}}(t) = k_{\text{fast}} \cdot \sigma_{\text{local}}(t), k_{\text{fast}} < k_{\text{slow}}, \quad (4)$$

where $k_{\text{slow}}$ corresponds to the slow scale encoding with lower triggering sensitivity, which is used to characterize the slowly varying trend components of the signal. In contrast, the smaller coefficient $k_{\text{fast}}$ enables the fast scale encoding to respond more sensitively to subtle variations in the residual signal.

Specifically, the encoding process first applies delta modulation to the original signal $x(t)$ under the lower sensitivity scale, generating the slow scale event sequence $s_{\text{slow}}(t)$ and its corresponding reconstruction signal $s_{\text{fast}}(t)$. This stage primarily extracts low-frequency trend information in sleep EEG and yields a relatively low event firing rate. Subsequently, the residual signal is computed as:

$$e(t) = x(t) - r_{\text{slow}}(t). \quad (5)$$

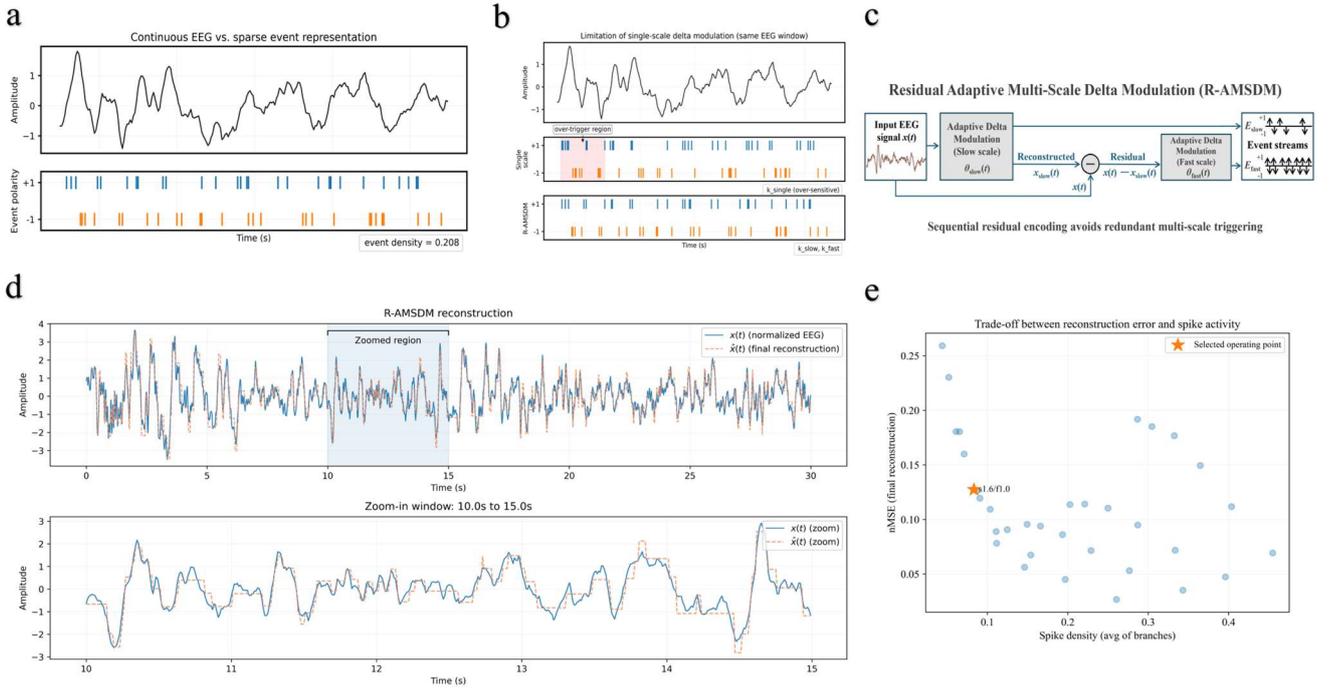

**Figure 2**. **R-AMSDM for neuromorphic EEG event encoding.** (**a**) Continuous EEG waveform and its sparse bipolar event representation; (**b**) Limitation of single-scale delta modulation and redundancy suppression by R-AMSDM; (**c**) Block diagram of the proposed R-AMSDM scheme; (**d**) EEG signal reconstruction from R-AMSDM events; (**e**) Trade-off between reconstruction error and event sparsity, with the selected operating point.

R-AMSDM then performs delta modulation on the residual signal $e(t)$ under the threshold $\theta_{\text{fas}}(t)$, producing the fast scale event sequence $s_{\text{fast}}(t)$. The final output of R AMSDM consists of both the slow scale event sequence $s_{\text{slow}}(t)$ and the fast scale event sequence $s_{\text{fast}}(t)$, which semantically correspond to the slowly varying physiological trends and transient fluctuations in sleep EEG, respectively.

This residual based multi-scale design explicitly suppresses cross scale redundant triggering while maintaining event sparsity, thereby enabling complementary dynamic representations across different temporal scales.

*2.2 Fidelity-Efficiency Trade-Off*

During the event-driven encoding stage, parameter settings directly determine the sparsity level of generated events, thereby influencing the computational load of subsequent event processing and temporal modeling. However, excessively sparse event representations may introduce nonnegligible information loss. Consequently, the selection of encoding parameters is formulated as a system-level trade-off between information fidelity and event sparsity [28, 29].

First, as illustrated in figure 2(d), a reconstruction consistency-based feasibility constraint is introduced to regulate event induced distortion without requiring additional supervision. The reconstructed signal is defined as:

$$\hat{x}(t) = r_{\text{fast}}(t) + r_{\text{slow}}(t), \tag{6}$$

where $r_{\text{slow}}(t)$ denotes the reconstruction component produced by slow scale delta modulation of the original signal, and $r_{\text{fast}}(t)$ represents the compensation term reconstructed from fast scale modulation of the residual signal. By evaluating the consistency between the original signal $x(t)$ and the reconstructed signal $\hat{x}(t)$, a set of feasibility constraints is constructed using the signal to noise ratio (SNR), normalized mean squared error (nMSE), and correlation coefficient, given by:

$$C = \{\text{SNR} \geq \tau_{\text{snr}}, \text{nMSE} \leq \tau_{\text{nmse}}, \text{Corr} \geq \tau_{\text{corr}}\}, \tag{7}$$

where $\tau_{\text{snr}}$, $\tau_{\text{nmse}}$, and $\tau_{\text{corr}}$ denote the acceptable thresholds for each metric. Only parameter configurations satisfying all constraints are regarded as feasible solutions, thereby bounding the distortion introduced by event encoding.

Under the information fidelity constraints, event density is further employed as a proxy metric for computational load during the encoding stage. Event density is defined as the proportion of time steps at which events are generated, and can be expressed as:

$$\rho = \frac{1}{T} \sum_{t=1}^{T} \mathbf{I}(s(t) \neq 0), \tag{8}$$

where $T$ denotes the encoding duration and $\mathbf{I}(\cdot)$ is the indicator function. Since computation triggering in event driven systems is highly correlated with the number of generated events, $\rho$ provides a reasonable estimate of relative energy consumption under different encoding parameter settings [30].

Based on this formulation, a grid search is performed over $k_{\text{slow}}$ and $k_{\text{fast}}$ to determine the operating point of the encoding stage. As shown in figure 2(e), event encoding is executed for each candidate parameter combination, and the corresponding reconstruction consistency metrics and event density $\rho$ are evaluated. Parameter configurations that violate the information fidelity constraints are first discarded. Among the remaining feasible solutions, the configuration yielding the minimum event density is selected as the final encoding setting, thereby minimizing the event triggering rate under the imposed constraints. Following this procedure, the parameters are set to $k_{\text{slow}} = 1.6$ and $k_{\text{fast}} = 1.0$.

*2.3 Signal-to-Event (S2E) Temporal Orchestration*

After neuromorphic encoding, the continuous EEG signal is transformed into an asynchronous event stream carrying timestamp and polarity information. To avoid temporal inconsistency and contextual mixing across discontinuous segments, a lightweight Signal to Event (S2E) temporal orchestration mechanism is introduced to perform event level time alignment and validity enforcement on the event stream.

Let the $n$-th event be denoted as $(t_n, p_n)$, where $t_n$ represents the triggering time of the event on the physical EEG time axis and $p_n \in \{+1, -1\}$ denotes the event polarity. S2E first assigns events to epochs using an epoch-level index and maintains the midpoint time of each epoch on the physical time axis. Let $T$ denote the epoch duration, which is set to 30 s. The epoch index and temporal anchor of each event are defined as:

$$e_{\text{idx}}(n) = \left\lfloor \frac{t_n}{T} \right\rfloor, t_{\text{mid}}(e) = \left(e + \frac{1}{2}\right)T, \tag{9}$$

where $t_{\text{mid}}$ serves as a temporal anchor that enables reliable characterization of the true temporal spacing between adjacent epochs, even when truncation or invalid segments exist in the event sequence.

Based on this formulation, an epoch-level validity mask $m_e \in \{0, 1\}$ is constructed to explicitly encode temporal continuity. When the time interval between adjacent epochs deviates from the expected step size $T$ by more than a tolerance $\tau = 0.1$ s, the interval is regarded as invalid or temporally discontinuous, and the corresponding position is marked as invalid, given by:

$$m_e = \mathbf{I}(|t_{\text{mid}}(e) - t_{\text{mid}}(e-1) - T| \leq \tau). \tag{10}$$

This validity mask is explicitly propagated within the model to constrain subsequent event-level modeling, thereby preventing contextual aggregation and state accumulation across discontinuous segments. Through this temporal orchestration strategy, the event-driven model maintains consistent and controllable temporal behavior on the physical time axis.

## 3. Neuromorphic event-based neural networks

### 3.1 Event-Based Adaptive Multi-Scale Response (EAMR)

Although S2E orchestration ensures temporal alignment and continuity of the event sequence, the resulting representation remains highly sparse and explicitly bipolar. In addition, discriminative cues in sleep EEG are distributed across both fast and slow temporal scales, which introduces a trade-off between representational capacity and computational efficiency for local modeling [31]. Motivated by this observation, NeuroSleep proposes the Event-based Adaptive Multi-scale Response (EAMR) module as an event-driven front-end that adaptively models multi-scale event responses within a local temporal window. An overview of the proposed NeuroSleep, is summarized in figure 3(a).

After S2E orchestration, the input is organized as events grouped by epoch. For a single epoch, the output of R-AMSDM consists of bipolar event sequences from both slow and fast scales, where event values belong to the set $\{-1, 0, +1\}$. These sequences are represented as a discrete signal $\mathbf{S} \in \mathbb{Z}^{2 \times T_b}$ with length $T_b$. Since the sign of each event explicitly encodes the direction of signal change, directly applying linear convolution within the same channel would weaken directional feature representation due to sign cancellation. To address this issue, EAMR first performs polarity expansion on the event sequence, which is illustrated in figure 3(b).

Specifically, for each event value $x$, positive and negative components are defined as $x^+ = \max(x, 0)$ and $x^- = \max(-x, 0)$, respectively. This yields a four-channel tensor corresponding to positive and negative events at slow and fast scales, given by:

$$\mathbf{E} = \Phi(\mathbf{S}) = [\mathbf{S}_{\text{slow}}^+, \mathbf{S}_{\text{slow}}^-, \mathbf{S}_{\text{fast}}^+, \mathbf{S}_{\text{fast}}^-] \in \mathbb{R}^{4 \times T_b}. \tag{11}$$

This expansion maps directional information from event polarity to channel representation without introducing additional temporal resolution, thereby avoiding feature attenuation caused by sign cancellation [32].

Based on this representation, EAMR employs parallel multi-scale temporal response branches to jointly capture local structures at different temporal extents. Specifically, three groups of one-dimensional depthwise separable convolutions with different receptive fields are adopted as the basic operators to reduce computational and parameter complexity. For a given scale $k \in \{k_s, k_m, k_l\}$, corresponding to small, medium, and large temporal scales, the branch output is expressed as:

$$\mathbf{H}_k = \sigma(\mathrm{BN}(\mathrm{PW}(\mathrm{DWConv}_k(\mathbf{E})))) \in \mathbb{R}^{w \times T_b}, \tag{12}$$

where $\mathrm{DWConv}_k(\cdot)$ denotes the one-dimensional depthwise convolution with kernel size $k$, $\mathrm{PW}(\cdot)$ is pointwise convolution for cross-channel fusion, BN represents batch normalization, and $\sigma(\cdot)$ indicates the GELU activation function to enhance nonlinear representation capacity. Responses from different temporal scales are subsequently concatenated along the channel dimension and projected to form a unified local feature representation $\mathbf{H} \in \mathbb{R}^{C \times T_b}$.

Considering that event firing rates vary across epochs and sleep stages, fixed-magnitude responses may introduce unnecessary interference in low-activity segments. To address this issue, EAMR incorporates a channel-wise adaptive gating mechanism that dynamically modulates the effective response magnitude of $\mathbf{H}$. Specifically, a channel descriptor is obtained via global aggregation of $\mathbf{H}$ along the temporal dimension, followed by a lightweight nonlinear mapping to produce a gating vector $\mathbf{g} \in (0, 1)^C$. The gated output is then given by:

$$\widetilde{\mathbf{H}} = \mathbf{H} \odot \mathbf{g}. \tag{13}$$

This channel-wise gating reflects the current level of event activity and enables adaptive response scaling, while maintaining a compact model structure. The resulting representation provides subsequent temporal modeling modules with locally adaptive features whose response strength is adjusted to event dynamics [33].

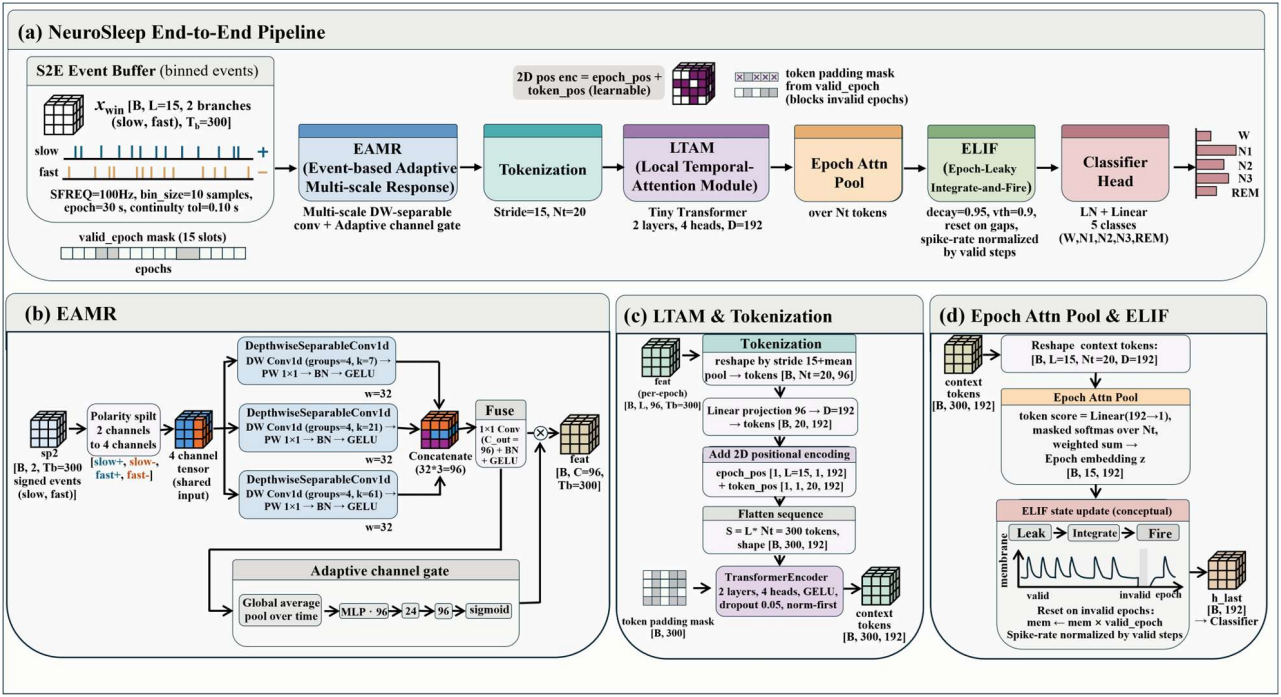

**Figure 3**. **The framework of the proposed NeuroSleep architecture and its key components.** (**a**) End-to-end NeuroSleep pipeline; (**b**) Structure of the EAMR module; (**c**) Tokenization process and LTAM; (**d**) Epoch-level attention pooling and ELIF state update.

*3.2 Local Temporal-Attention Module (LTAM)*

Through the EAMR front-end, the NeuroSleep obtains a polarized multi-scale local representation of sparse events within a single epoch. However, this representation remains confined to the temporal scope of an individual epoch and does not explicitly characterize the evolutionary relationships across adjacent epochs [34].

For sleep staging tasks, stage discrimination often relies on gradual transitions spanning multiple neighboring epochs rather than isolated local structures. Therefore, while preserving the event-driven computation paradigm and maintaining controlled computational cost, NeuroSleep introduces a limited temporal context modeling mechanism and proposes the Local Temporal-Attention Module (LTAM). The detailed structure is shown in figure 3(c).

Specifically, LTAM takes the epoch-level feature sequence $\{\widetilde{\mathbf{H}}_i\}_{i=1}^{N}$ produced by EAMR as input, where each element corresponds to a local representation of an epoch. For a given epoch $i$, LTAM aggregates contextual information only within a local temporal neighborhood. This neighborhood is constrained by a temporal window radius $L$ and the validity mask $m_e$ generated during the S2E stage, and is encoded by an attention mask defined as:

$$M_{i,j} = \begin{cases} 0, & \text{if } |i-j| \leq L \land m_e(j) = 1 \\ -\infty, & \text{otherwise.} \end{cases} \tag{14}$$

Within the local window permitted by this mask, LTAM employs a scaled dot-product attention mechanism to selectively aggregate features from neighboring epochs.

Specifically, for the central epoch $i$, its feature representation is projected into a query vector $\mathbf{q}_i$, while the features of visible neighboring epochs are projected into key vectors $\mathbf{k}_j$ and value vectors $\mathbf{v}_j$. The attention weights and contextual aggregation are jointly expressed as:

$$\alpha_{i,j} = \text{softmax}_j\left(\frac{\mathbf{q}_i^\top \mathbf{k}_j}{\sqrt{d}} + M_{i,j}\right), \mathbf{z}_i = \sum_j \alpha_{i,j} \mathbf{v}_j, \tag{15}$$

where $d$ denotes the feature dimensionality, and the softmax operation is normalized only over the visible epochs. In contrast to global attention mechanisms designed for long sequence modeling, LTAM restricts attention computation to a bounded local temporal window. This design enhances discriminative information around stage transition regions while maintaining low computational complexity and preserving the efficiency of event-driven modeling [35].

### 3.3 Epoch-Leaky Integrate-and-Fire (ELIF) State Module

After completing local event representation with EAMR and local cross-epoch association with LTAM, NeuroSleep obtains context-aware epoch representations. However, sleep stages exhibit state persistence over longer time scales. Relying solely on associations within a limited local window is insufficient to provide stable memory across epochs. To address this limitation, NeuroSleep introduces a biologically inspired Epoch-Leaky Integrate-and-Fire (ELIF) state module, as shown in figure 3(d).

ELIF formulates cross-epoch temporal modeling as a leaky integration process operating at a low temporal frequency. Its design is inspired by the accumulation and decay dynamics of biological neurons over extended time scales [36]. Let $\mathbf{z}_i$ denote the feature representation of the $i$-th epoch produced by LTAM. ELIF maintains a cross-epoch state variable $\mathbf{h}_i$, which is updated under temporal continuity according to the following leaky integration rule:

$$\mathbf{h}_i = \lambda \mathbf{h}_{i-1} + \mathbf{z}_i, \tag{16}$$

where $\lambda \in (0, 1)$ is the leakage coefficient that controls the influence of historical states on the current representation. This update mechanism allows short-term fluctuations to gradually decay, while discriminative features that persist across stable sleep stages are accumulated in the state variable, thereby naturally reflecting stage continuity. Notably, when S2E indicates a temporal discontinuity, the state is explicitly reset to prevent contamination across discontinuous segments and to preserve physical time consistency.

In addition, the number of valid epochs may vary across samples, which can introduce magnitude bias in the accumulated state. To mitigate this effect, ELIF applies normalization based on the number of effective updates. Let $n_i$ represent the number of valid state updates up to the $i$-th epoch. The normalized state is defined as:

$$\bar{\mathbf{h}}_i = \frac{\mathbf{h}_i}{\max(n_i, 1)}. \tag{17}$$

This normalization encourages the state representation to emphasize long-term trends rather than simple accumulation magnitude, thereby yielding more stable discrimination near stage boundaries. Finally, the normalized state representation is provided as a low-frequency abstract feature to a lightweight classifier for sleep stage prediction, forming a coherent system-level information flow together with the neuromorphic event encoding and local temporal modeling modules [37].

## 4. Experimental setup and evaluation

*4.1 Datasets and model training*

Experiments were conducted on the sleep-cassette (SC) subset of the Sleep-EDF Expanded database, using single-channel EEG recorded from Fpz-Cz as input [38]. The SC subset comprises 153 overnight recordings from 78 subjects. The raw signals were uniformly band-pass filtered between 0.5 and 35 Hz and resampled to 100 Hz. The processed signals were then segmented into 30 s epochs and annotated into five sleep stages according to the AASM standard, namely W, N1, N2, N3, and REM [39].

A subject-independent 5-fold cross-validation protocol was adopted and the 78 subjects were partitioned into five folds, ensuring that test subjects were strictly separated from the training and validation cohorts to evaluate the generalizability of system to unseen individuals. Within each training fold, 15% of the subjects were further held out as a validation set. A local temporal context window of length $L$=15 was constructed centered on the target epoch. Training was performed using AdamW with a learning rate of $1\times10^{-3}$ and a weight decay of $1\times10^{-4}$, a batch size of 64, and a maximum of 50 epochs. Early stopping with a patience of 8 epochs was applied, and automatic mixed precision training was enabled.

*4.2 Comparison and ablation experiment setup*

To evaluate the overall performance of NeuroSleep in terms of classification accuracy and computational efficiency, both comparative experiments (Group B) and ablation studies (Group A) were conducted. NeuroSleep represents the complete model, which integrates R-AMSDM, EAMR, LTAM, and the ELIF module.

The comparative models were designed to cover representative modeling paradigms. Specifically, B1 is an epoch-level convolutional network that performs modeling only within individual epochs. B2 is a dense token-based Transformer that retains the contextual window and tokenized input representation while operating under dense computation. B3 is a time-unrolled Spiking CNN baseline, in which Leaky Integrate-and-Fire (LIF) dynamics are applied at finer-grained time steps with step-by-step updates. B4 follows an AttnSleep-lite style architecture, combining convolutional feature extraction with lightweight attention aggregation, and serves as a strong representative baseline [40].

In parallel, a series of ablation experiments were conducted to quantify the contribution of key components in NeuroSleep. Accordingly, A1 removes R-AMSDM and degrades the input representation to dense continuous signals. A2 simplifies EAMR by reducing the parallel multi-scale responses to a single branch. A3 replaces LTAM with a lightweight temporal mixing module. A4 removes the ELIF state module entirely.

During evaluation, all models were assessed using a unified set of metrics, including classification accuracy and system-level computational indicators. These indicators comprise the number of model parameters, inference latency (ms/sample), dense floating-point operations (FLOPs), sparsity-adjusted effective operations (Effective OPs), input event density (inSD), and the ELIF state update rate (Spike Rate). Notably, Effective OPs account only for computation components that scale with event sparsity and are adjusted according to the inSD, which are then combined with computation that must be executed densely:

$$\text{Effective OPs} = \text{FLOPs}_{\text{sparse}} \times \text{inSD} + \text{FLOPs}_{\text{dense}}. \tag{18}$$

The ELIF Spike Rate reflects the average firing probability of the state module over valid time steps.

*4.3 Overall performance and efficiency*

To ensure the generalizability of NeuroSleep across different individuals, all reported metrics are the mean results obtained from 5-fold subject-independent cross-validation. As shown in table 1, NeuroSleep achieves the best discriminative performance among all evaluated models. It attains an Accuracy of 74.2%, a Macro-F1 score of 70.5%, and a Cohen's Kappa of 65.6%, all of which rank highest across the compared methods. From an engineering perspective, NeuroSleep contains 0.932 M parameters and achieves a single-sample inference latency of 1.047 ms. Under the event-driven execution setting, the inSD is 0.463, corresponding to 0.357 G Effective OPs. This represents a reduction of approximately 53.6% relative to the dense execution upper bound of 0.770 G. Meanwhile, the ELIF Spike Rate is 0.566, indicating a stable balance between energy suppression and state expressiveness. Figure 4 further summarizes the accuracy–efficiency trade-off, where NeuroSleep lies on the Pareto frontier.

| Model ID | Accuracy (↑) | Macro-F1 (↑) | Kappa (↑) | Params (M) (↓) | Latency (ms) (↓) | FLOPs (G) (↓) | Effective OPs (G) (↓) | inSD | Spike Rate |
|---|---|---|---|---|---|---|---|---|---|
| **NeuroSleep** | **74.2%** | **70.5%** | **65.6%** | 0.932 | 1.047 | 0.770 | 0.357 | 0.463 | 0.566 |
| A1 | 72.0% | 68.2% | 62.1% | 0.932 | 1.022 | 0.770 | 0.770 | 1.000 | 0.712 |
| A2 | 71.8% | 67.4% | 62.2% | 0.925 | 0.737 | 0.710 | 0.332 | 0.467 | 0.251 |
| A3 | 72.7% | 69.3% | 63.8% | 0.056 | 8.879 | 0.015 | 0.007 | 0.465 | 0.351 |
| A4 | 70.2% | 67.8% | 61.3% | 0.932 | 1.011 | 0.770 | 0.353 | 0.458 | N/A |
| B1 | 64.8% | 62.7% | 54.4% | 0.098 | 0.308 | 0.036 | 0.036 | N/A | N/A |
| B2 | 66.7% | 61.7% | 55.6% | 0.819 | **0.256** | 0.659 | 0.659 | N/A | N/A |
| B3 | 44.5% | 41.1% | 27.9% | **0.037** | 0.795 | **0.003** | **0.003** | N/A | N/A |
| B4 | 65.6% | 59.1% | 54.1% | 0.057 | 0.457 | 0.010 | 0.010 | N/A | N/A |

**Table 1**. Performance and efficiency comparison of NeuroSleep and ablation variants (mean over 5-fold subject-independent cross-validation).

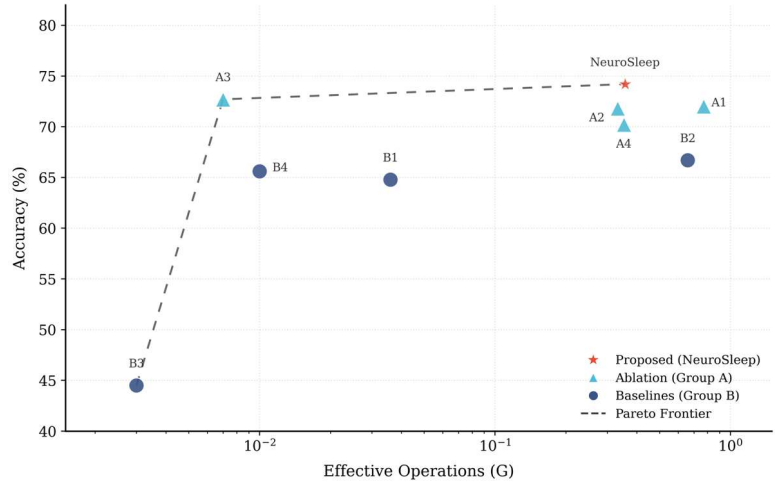

**Figure 4**. Accuracy–efficiency trade-off across evaluated models.

| Stage | Precision | Recall | F1-score |
|---|---|---|---|
| W | 92.7 ± 2.4% | 82.6 ± 2.4% | 87.3 ± 2.3% |
| N1 | 39.9 ± 3.9% | 52.8 ± 5.3% | 45.3 ± 3.1% |
| N2 | 83.6 ± 3.2% | 73.1 ± 3.1% | 77.9 ± 1.4% |
| N3 | 63.6 ± 8.1% | 81.5 ± 4.2% | 71.0 ± 5.3% |
| REM | 67.1 ± 2.5% | 75.4 ± 6.7% | 70.9 ± 3.4% |
| Macro Average | 69.4 ± 2.8% | 73.1 ± 2.9% | 70.5 ± 1.9% |

**Table 2**. Per-class precision, recall, and F1-score of NeuroSleep (mean ± standard derivation over 5-fold subject-independent cross-validation).

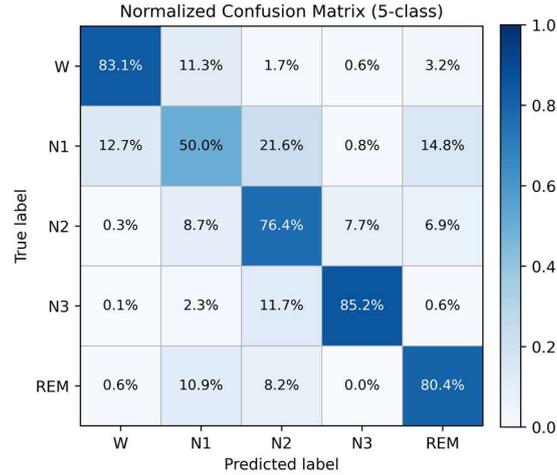

**Figure 5**. The confusion matrix of the NeuroSleep for five-class sleep staging (W, N1, N2, N3, REM).

As further illustrated in table 2 and figure 5, NeuroSleep exhibits a largely diagonal confusion pattern under the five-class setting, indicating stable stage-wise discrimination. High recall is achieved for relatively stable stages (W, N2, N3, and REM), whereas the main errors concentrate on N1, which is primarily confused with adjacent stages, particularly N2 and REM.

Compared with representative baselines, NeuroSleep demonstrates a clear advantage in the joint trade-off between performance and efficiency. B1, which lacks cross-epoch contextual modeling, achieves an Accuracy of only 64.8%. B2, which incorporates dense token representations with a contextual window, improves performance to 66.7% but remains 7.5% lower in Accuracy than NeuroSleep. Moreover, its Effective OPs remain at 0.659 G, failing to benefit from sparse execution, relative to which NeuroSleep achieves a 45.8% reduction in computational load. Although B3 exhibits very low FLOPs, it delivers the weakest performance with an Accuracy of 44.5%, indicating that naive time-step-based spiking neural networks are not well suited for efficient sleep staging. B4 serves as a strong baseline with relatively low complexity, yet its performance remains limited, achieving an Accuracy of 65.6%.

Ablation studies further validate the necessity of each module within NeuroSleep. Removing R-AMSDM in A1 reduces Accuracy to 72.0%, while Effective OPs increase to 0.770 G and the Spike Rate rises to 0.712. This result indicates that event encoding not only provides sparsity-induced computational benefits but also suppresses redundant activations and stabilizes state updates. Simplifying EAMR to a single-scale response in A2 leads to a further performance drop, with Accuracy decreasing to 71.8% and the Spike Rate falling to 0.251. This suggests that multi-scale event responses are critical for maintaining effective driving signals and discriminative capacity. Replacing LTAM with a lightweight temporal mixing module in A3 preserves a comparable Accuracy of 72.7%, but inference latency increases significantly to 8.879 ms, revealing the efficiency bottleneck of sequential or poorly parallelizable structures in edge deployment. Finally, removing the ELIF state module in A4 further degrades Accuracy to 70.2%, highlighting the importance of low-power cross-epoch state memory for stable sleep stage prediction.

In summary, the advantages of NeuroSleep do not stem from any single component, but rather from the system-level synergy of event-driven encoding, multi-scale response, local attention, and state memory. This coordinated design substantially reduces sparsity-adjusted effective computation without sacrificing classification accuracy, thereby providing an effective solution for real-time sleep staging in edge computing scenarios.

**Discussion**

This paper proposes NeuroSleep, an event-driven modeling paradigm for sleep staging targeting sensor-level and edge-level applications. Unlike conventional dense inference schemes that rely on continuous high-frequency signals, NeuroSleep employs R-AMSDM to convert raw EEG into a change-driven sparse event stream, thereby suppressing redundant processing originating from temporally stationary segments at the input stage. As a result, the proposed framework significantly reduces effective computation and state update frequency during edge inference, while preserving discriminative performance, achieving a more suitable balance between accuracy and efficiency for resource-constrained devices.

Furthermore, the hierarchical modeling strategy of NeuroSleep elucidates the role of event-driven approaches in physiological signal analysis. Through multi-scale response modeling and limited temporal context selection, EAMR and LTAM enable robust and stable sleep staging under low computational complexity. The ELIF state module further reinforces stage persistence via low-frequency state memory, while avoiding the efficiency bottlenecks commonly observed in SNN deployed at the edge. Together, these components form an end-to-end inference architecture that is well aligned with event sparsity. Nevertheless, the current study focuses primarily on single-channel EEG, leaving room for further exploration in multi-modal sensing and cross-dataset generalization. In addition, the stability of event density and sparsity-related gains across different subjects and acquisition conditions warrants systematic evaluation in broader experimental settings.

## Conclusion

This paper presents NeuroSleep, an event-driven single-channel EEG sleep staging framework designed for sensor-level and edge-level applications. Centered on a change-driven principle, the proposed approach converts continuous high-frequency EEG signals into sparse event representations through neuromorphic event encoding, and combines this representation with hierarchical temporal modeling to effectively characterize sleep stage dynamics. In contrast to conventional methods that rely on continuous dense inference, NeuroSleep reconstructs the processing pipeline from signal acquisition and representation to inference at the system-level. This design enables a substantial reduction in effective computational load and state update frequency during edge inference, while maintaining reliable sleep staging performance, thereby offering a practical design paradigm for wearable and long-term edge-based sleep analysis.

Beyond the specific model architecture, this study introduces a generalizable event-driven paradigm for physiological signal analysis. By explicitly separating high temporal resolution change detection from low-frequency abstract context modeling, NeuroSleep provides a new perspective for efficient on-device processing of complex temporal signals. Future work will focus on multi-modal sensory fusion, cross-dataset and cross-device generalization, and the co-optimization of event statistics and model structure, in order to further validate the applicability and scalability of this paradigm across broader physiological monitoring scenarios.

## Acknowledgements

The authors would like to thank the contributors of the Sleep-EDF Expanded database available on PhysioNet for providing the open-access sleep EEG data used in this study.

## Data availability statement

The data that support the findings of this study are openly available in the Sleep-EDF Database Expanded (v1.0.0) at PhysioNet (https://doi.org/10.13026/C2X676).

## Ethical statement

All human EEG data used in this study were obtained from the publicly available Sleep-EDF Expanded database on PhysioNet. The original data collection was conducted in accordance with the Declaration of Helsinki and was approved by the local ethics committees of the respective institutions where the data were recorded.

Physiological Measurement